\documentclass[11pt,amsmath,amssymb,noeprint]{revtex4-1}

\usepackage{amsfonts}
\usepackage{amsmath}
\usepackage{amssymb}
\usepackage{array}
\usepackage{bbding}
\usepackage{bm}
\usepackage{booktabs}
\usepackage{braket}
\usepackage{breqn}
\usepackage{dcolumn}
\usepackage{diagbox}
\usepackage{endnotes}
\usepackage{epsf}
\usepackage{epsfig}
\usepackage{epstopdf}
\usepackage{extarrows}
\usepackage{float}
\usepackage{graphicx}
\usepackage{harpoon}
\usepackage{hyperref}
\usepackage{indentfirst}
\usepackage{makecell}
\usepackage{mathrsfs}
\usepackage{mathtools}
\usepackage{multirow}
\usepackage{pifont}
\usepackage{slashed}
\usepackage{xcolor}

\newcommand{\beq}{\begin{eqnarray}}
\newcommand{\eeq}{\end{eqnarray}}

\begin{document}

\title{$\gamma Z$-exchange contribution in elastic $ep$ scattering by perturbative QCD}

\author{
Qian-Qian Guo$^{1,2}$\footnotemark[1]\footnotetext[1]{E-mail: guoqq@lit.edu.cn},
Hui-Yun Cao$^{3}$,
and Hai-Qing Zhou$^{2}$\footnotemark[2]\footnotetext[2]{E-mail: zhouhq@seu.edu.cn} \\
$^1$Department of Mathematics and Physics, Luoyang Institute of Science and Technology, Luoyang 471023, China\\
$^2$School of Physics, Southeast University, Nanjing 211189, China\\
$^3$ School of Physics and Electronic Science, Hubei Normal University, Huangshi 435002, China\\
}

\date{\today}

\begin{abstract}
In this study, we calculate the $\gamma Z$-exchange contribution to elastic $ep$ scattering at large momentum transfer within perturbative QCD. We present analytical expressions for the $\gamma Z$-exchange contributions to the amplitudes. We also estimate the asymptotic behaviors of the amplitude contributions and of the physical quantity $A_{\text{PV}}$ at high momentum transfer. These asymptotic behaviors determine the subtraction order in the dispersion relations (DRs) satisfied by the amplitudes. \textcolor{black}{At high $Q^2$ beyond the forward limit, our pQCD analysis shows that this subtraction order is fixed by the high-$\nu$ behavior of the amplitude coefficients $c_i^{\gamma Z,\text{PV}}$, and that $c_{3,A}^{\gamma Z,\text{PV}}$ requires a once-subtracted DR.} Within the present pQCD framework and the adopted proton distribution amplitudes, these high-energy results also provide nontrivial constraints on low-energy DR assumptions.
\end{abstract}

\maketitle

\section{Introduction}
The two-photon-exchange (TPE) contribution has been shown to have a sizable effect on the extraction of electromagnetic form factors (FFs) in unpolarized $ep$ scattering. Similarly, the $\gamma Z$-exchange contribution also plays an important role in extracting the proton weak charge and strange FFs. Many methods have been applied to estimate these two-boson-exchange (TBE) contributions, including hadronic models \cite{hadron-model-Zhou:2007hr,hadron-model-Nagata:2008uv,hadron-model-Tjon:2007wx,hadron-model-Tjon:2009hf}, general parton distributions \cite{GPDs-Chen:2009mza,GPDs-Chen:2004tw,GPDs-Afanasev:2005ex}, perturbative QCD (pQCD) \cite{pQCD-Borisyuk:2009eg,pQCD-Kivel:2009eg,pQCD-Guttmann:2011ze---ppbar-ee}, DR methods \cite{DRs-Gorchtein:2008px,DRs-Sibirtsev:2010zg,DRs-Gorchtein:2011mz,DRs-Erler:2019rmr}, and lattice QCD \cite{LQCD-Fu:2022fgh,LQCD-Xiong:2023zih,LQCD-Pachucki:2022tgl,LQCD-Salg:2025cgu,LQCD-Zhang:2026utt}. Among these methods, pQCD provides a reliable prediction of the high-energy behavior of these contributions. These behaviors are directly useful at high energy and also indicate the subtraction properties in the DR method. The TPE contribution in elastic $ep$ scattering has been studied in Ref.~\cite{pQCD-Kivel:2009eg}, \textcolor{black}{and the corresponding $\gamma Z$-exchange contribution in pQCD was investigated in Ref.~\cite{Guttmann:2013ndr}. We extend those studies by presenting the results directly in terms of the invariant amplitudes $c_i^{\gamma Z,\text{PV}}$, including the full set of parity-violating coefficients and their high-$\nu$ behavior at large $Q^2$. For the DR analysis, we focus on these amplitude coefficients rather than on the combinations $\square_{\gamma Z}^{V,A}$ that enter $A_{\text{PV}}$. As shown in Ref.~\cite{QianQianGuo2024}, DRs written directly for $\square_{\gamma Z}^{V,A}$ are reliable only in the forward approximation, whereas DRs for $c_i^{\gamma Z,\text{PV}}$ or for the helicity amplitudes provide the appropriate starting point at finite $Q^2$.}

In this work, we extend the pQCD calculation of the TPE contribution in $ep$ scattering to the $\gamma Z$-exchange contribution at large $Q^2$, and we analyze the high-$\nu$ behavior of the amplitude coefficients $c_i^{\gamma Z,\text{PV}}$ and the implications for their DRs beyond the forward limit. In Sec.~II, we present the basic formalism, including the amplitude decomposition, the relevant couplings, and the pQCD setup for TPE and $\gamma Z$ exchange. In Sec.~III, we give the analytical results for the invariant amplitudes. In Sec.~IV, we present the numerical analysis, extract the high-energy asymptotic behaviors, and discuss the implications for DRs of $c_i^{\gamma Z,\text{PV}}$. Sec.~V summarizes the main conclusions.

\section{Basic Formalism}

\subsection{Expressions for two-photon-exchange contribution in elastic $ep$ within pQCD}
The TPE contribution in elastic $ep$ scattering is discussed in Ref.~\cite{pQCD-Kivel:2009eg} using the pQCD method. Here we review the basic formalism used in this method. Within pQCD for $ep$ scattering, at leading order in the coupling, there are 24 Feynman diagrams, and some of them are shown in Fig.~1. The momenta of the incoming electron, outgoing electron, incoming proton, and outgoing proton are labeled as $p_{1,3,2,4}$, respectively. $m_e$ is the electron mass and $m_N$ is the proton mass. For convenience, we define
\begin{eqnarray}
&P\equiv\frac{1}{2}(p_2+p_4), ~~K\equiv\frac{1}{2}(p_1+p_3),~~q\equiv p_1-p_3, \nonumber\\
&Q^2\equiv -q^2,~~\nu\equiv K \cdot P,~~s\equiv(p_1+p_2)^2.
\end{eqnarray}

\begin{figure}[!h]
\begin{center}
\includegraphics[height=5.5cm]{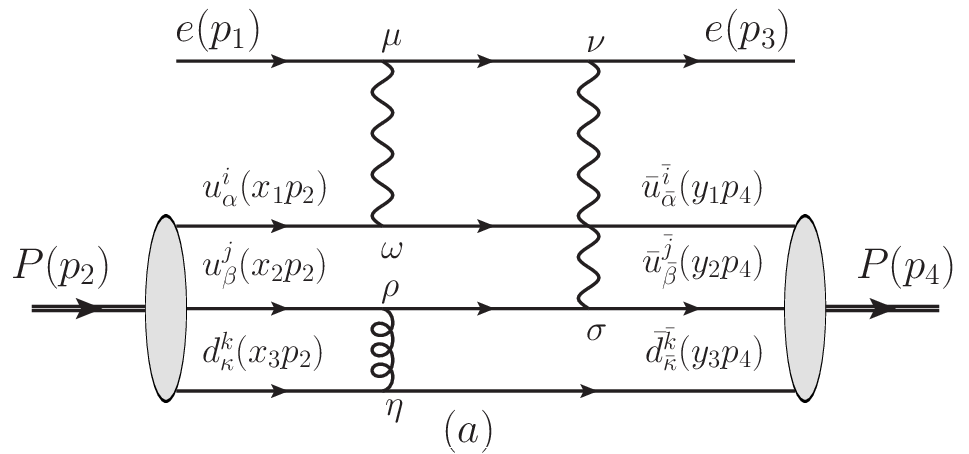}
\caption{One of the Feynman diagrams for the TPE contribution in $ep\rightarrow ep$, where the two photons are connected to two $u$ quarks, and the gluon is connected to the $u$ and $d$ quarks.}
\label{Figure:ep-ep-TPE-uu-a}
\end{center}
\end{figure}
The corresponding amplitude for the Feynman diagram in Fig.~\ref{Figure:ep-ep-TPE-uu-a} is
\begin{eqnarray}
\mathcal{M}_{2\gamma}^{(a)} &=& -i \int^{1}_{0} [dxdy] \bar{u}_{3} \Gamma_{\gamma ee}^{\nu} S_e(p_1-y_1p_4+x_1p_2) \Gamma_{\gamma ee}^{\mu} u_{1} \Phi_{P}^{\text{fin}}(\bar{i},\bar{j},\bar{k},\bar{\alpha},\bar{\beta},\bar{\kappa},p_4, \lambda_{4}) [\Gamma_{\gamma uu}^{\omega}]_{\bar{\alpha}\alpha}\delta_{\bar{i}i} \nonumber\\
~~~~~~&&\times[\Gamma_{\gamma uu}^{\sigma}S_q((x _2+x_3)p_2-y_3p_4)\Gamma_{guu}^{\rho}]_{\bar{\beta}\beta}T^a_{\bar{j}j}[\Gamma_{gdd}^{\eta}]_{\bar{\kappa}\kappa}
T^a_{\bar{k}k}\Phi_{P}^{\text{ini}}(i,j,k,\alpha,\beta,\kappa,p_2, \lambda_{2}) \nonumber\\
~~~~~~&& \times D_{g,\rho\eta}[y_3p_4-x_3p_2]D_{\gamma,\mu\omega}(y_1p_4-x_1p_2)D_{\gamma,\nu\sigma}((y_2+y_3)p_4-(x_2+x_3)p_2),
\label{eq:Amp-2gamma-uu-a}
\end{eqnarray}
where $x_{1,2,3},y_{1,2,3}$ are the corresponding momentum factors of the quarks in the proton and
\begin{eqnarray}
[dxdy]\equiv dx_1dx_2dx_3dy_1dy_2dy_3\delta(x_1+x_2+x_3-1)\delta(y_1+y_2+y_3-1),
\end{eqnarray}
with
\begin{eqnarray}
x_1+x_2+x_3&=&1,  ~~~~0\leq x_l\leq 1,  \nonumber\\
y_1+y_2+y_3&=&1,  ~~~~ 0\leq y_l\leq 1.
\end{eqnarray}
The indices $i,j,k$ and $\bar{i},\bar{j},\bar{k}$ denote quark color, $\alpha,\beta,\kappa$ and $\bar{\alpha},\bar{\beta},\bar{\kappa}$ are Dirac indices, $a,b$ are gluon color indices, $\lambda_{2,4}$ are the helicities of the incoming and outgoing proton, and $\Phi_{\text{P}}^{\text{fin/ini}}$ are the distribution amplitudes of the outgoing and incoming protons in pQCD. The propagators $S_e(k)$, $D_{\gamma,\mu\nu}(k)$, $S_q(k)$, and $D_{g,\mu\nu}(k)$ for the electron, photon, quark, and gluon are
\begin{eqnarray}
D_{\gamma,\mu\nu}(k) &=& \frac{-i}{k^2+i\varepsilon}g_{\mu\nu}, \quad S_e(k) = \frac{i(\slashed{k}+m_e)}{k^2-m_e^2+i\varepsilon} \nonumber \\
D_{g,\mu\nu}(k) &=& \frac{-i}{k^2+i\varepsilon}g_{\mu\nu}, \quad S_q(k) = \frac{i\slashed{k}}{k^2-m_q^2+i\varepsilon},
\label{eq:propagators}
\end{eqnarray}
The vertices are
\begin{eqnarray}
\Gamma_{\gamma ee}^{\mu} &=& -ie\gamma^{\mu}, \quad \Gamma_{\gamma uu}^{\mu} = iQ_ue\gamma^{\mu} \nonumber \\
\Gamma_{\gamma dd}^{\mu} &=& iQ_de\gamma^{\mu}, \quad \Gamma_{gdd/uu}^{\mu} = -ig_s\gamma^{\mu},
\label{eq:vertices-gamma-g}
\end{eqnarray}
where \textcolor{black}{$e$ is the electromagnetic interaction coupling constant, $g_s$ is the strong interaction coupling constant and} $Q_u=2/3$ and $Q_d=-1/3$ are the quark charges.  The distribution amplitudes
are expressed as:
\begin{eqnarray}
\Phi_{P}^{\text{ini}}[i,j,k,\alpha,\beta,\kappa,p,\lambda] &=&\frac{1}{24}\epsilon_{ijk} \Big\{ V [\slashed{p} C]_{\alpha \beta}[\gamma_5 u(p,\lambda)]_\kappa +A [\slashed{p} \gamma_5 C]_{\alpha \beta}[u(p,\lambda)]_\kappa \nonumber \\
&&~~~~~~~~~~+T [\sigma_{\mu\nu}p^{\nu} C]_{\alpha \beta}[\gamma^{\mu} \gamma_5 u(p,\lambda)]_\kappa \Big \},\nonumber\\
\Phi_{P}^{\text{fin}}[i,j,k,\alpha,\beta,\kappa,p,\lambda] &=& (\Phi_{P}^{\text{ini}}[i,j,k,\alpha,\beta,\kappa,p,\lambda] )^{\ast},
\end{eqnarray}
where $C$ is the charge-conjugation matrix, $u^{+}(p,\lambda)$ is the plus component of the Dirac spinor, and the functions $A,V,T$ are scalar functions. The above distribution amplitudes correspond to the matrix element in Ref.~\cite{pQCD-Kivel:2009eg}:
\begin{eqnarray}
\epsilon_{ijk}\Phi_{P}^{\text{ini}}[i,j,k,\alpha,\beta,\gamma,p,\lambda] &\sim&4\left\langle 0\left| \epsilon_{ijk}{u}_\alpha^i\left(x_1 n\right) {u}_{\beta}^j\left(x_2 n\right) {d}_\kappa^k\left(x_3  n\right)\right| p,\lambda\right\rangle\nonumber\\
&=& V p^{+}\left[\left(\frac{1}{2} \bar{n} \cdot \gamma\right) C\right]_{\alpha \beta}\left[\gamma_5 N^{+}\right]_\sigma +A p^{+}\left[\left(\frac{1}{2} \bar{n} \cdot \gamma\right) \gamma_5 C\right]_{\alpha \beta}\left[N^{+}\right]_\sigma\nonumber\\
&& +T p^{+}\left[\frac{1}{2} i \sigma_{\perp \bar{n}} C\right]_{\alpha \beta}\left[\gamma^{\perp} \gamma_5 N^{+}\right]_\sigma,
\label{eq:Amp-DAs}
\end{eqnarray}
where the incoming-proton momentum satisfies $\frac{1}{2}\bar{n}p^+\sim p$ and the spinor satisfies $N^+ \sim u^+(p,\lambda)$. In the practical calculation, we take the functions $A,V,T$ to be the same as those used in Ref.~\cite{pQCD-Kivel:2009eg}:
\begin{eqnarray}
V(x_i) &=& 120x_1x_2x_3f_N[1+r_+(1-3x_3)], \nonumber \\
A(x_i) &=& 120x_1x_2x_3f_Nr_-(x_2-x_1), \nonumber \\
T(x_i) &=& 120x_1x_2x_3f_N[1+1/2(r_-r_+)(1-3x_3)].
\label{eq:T-A-V}
\end{eqnarray}
The parameters $f_N, r_{\pm}$ are listed in Tab.~\ref{tab:fN}, where three models are used: COZ \cite{COZ-definition}, BLW \cite{BLW-definition}, and QCDSF \cite{QCDSF-definition}.

\begin{table}[]
\centering
\begin{tabular}{|c|c|c|c|}
  \hline
      & $f_{N}(10^{-3})$ GeV$^2$ & $r_{-}$ & $r_{+}$ \\
  \hline
  COZ & 5.0 $\pm$ 0.5 & 4.0 $\pm$ 1.5 & 1.1 $\pm$ 0.3 \\
  \hline
  BLW & 5.0 $\pm$ 0.5 & 1.37 & 0.35 \\
  \hline
  QCDSF & 3.23 $\pm$ 0.06 $\pm$ 0.09 & 1.06 $\pm$ 0.09 $\pm$ 0.31 & 0.33 $\pm$ 0.03 $\pm$ 0.11 \\
  \hline
\end{tabular}
\caption{Parameters $f_N$ and $r_{\pm}$ in the COZ \cite{COZ-definition}, BLW \cite{BLW-definition}, and QCDSF \cite{QCDSF-definition} models.}
\label{tab:fN}
\end{table}

For the running coupling constant $\alpha_{s}(Q^2)\equiv g_s^2/(4\pi)$, we use
\begin{eqnarray}
\alpha_{s}(Q^2) &=& \frac{1}{b_0t}(1-\frac{b_1}{b_0^2}\frac{l}{t}),
\label{eq:Amp-alphas}
\end{eqnarray}
where
\begin{eqnarray}
l &=& \log[t], \quad b_0 = \frac{33-2n_f}{12\pi},\nonumber \\
t &=& \log[\frac{Q^2}{\Lambda_{\text{QCD}}^2}], \quad b_1 = \frac{153-19n_f}{24\pi^2}.
\label{eq:Amp-alphas-b0-b1}
\end{eqnarray}
with $\Lambda_{\text{QCD}}=0.2$ GeV and $n_f=4$.

\textcolor{black}{We note that, at the leading order considered here ($\alpha_e^2\alpha_s$), the box diagrams in Figs.~\ref{Figure:ep-ep-TPE-uu-a} and~\ref{Figure:ep-ep-gammaZ-uu-1-a} contain no explicit infrared divergence. Infrared divergences from additional gluon exchanges appear only at the next order ($\alpha_e^2\alpha_s^2$), which is beyond the scope of the present work.}

\subsection{$\gamma Z$-exchange contribution in elastic $ep$ scattering}
\begin{figure}[!h]
\begin{center}
\includegraphics[height=5.5cm]{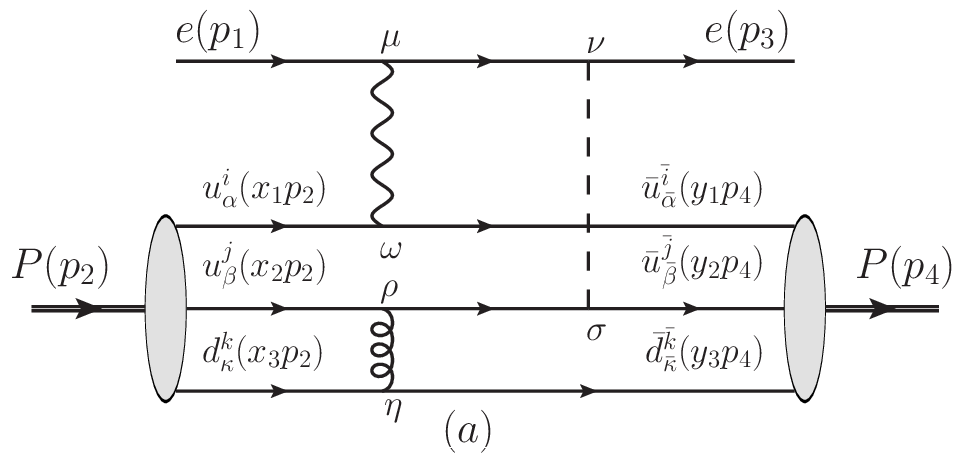}
\caption{One of the Feynman diagrams for $\gamma Z$ exchange in $ep\rightarrow ep$, where the photon and the $Z$ boson are attached to two $u$ quarks, and the gluon is attached to the $u$ and $d$ quarks.}
\label{Figure:ep-ep-gammaZ-uu-1-a}
\end{center}
\end{figure}
Similar to the TPE case, the $\gamma Z$-exchange contribution can also be estimated by pQCD at high energy, and there are 48 Feynman diagrams for this process at leading order. We show one diagram in Fig.~\ref{Figure:ep-ep-gammaZ-uu-1-a}, where the notations are the same as those used in the TPE case. The corresponding amplitude is
\begin{eqnarray}
\mathcal{M}_{\gamma Z}^{(a)} &=& -i \int^{1}_{0} [dxdy] \bar{u}_{3} \Gamma_{Zee}^{\nu} S_e(p_1-y_1p_4+x_1p_2) \Gamma_{\gamma ee}^{\mu} u_{1} \Phi_{P}^{\text{fin}}(\bar{i},\bar{j},\bar{k},\bar{\alpha},\bar{\beta},\bar{\kappa},p_4, \lambda_{4}) [\Gamma_{\gamma uu}^{\omega}]_{\bar{\alpha}\alpha}\delta_{\bar{i}i} \nonumber\\
~~~~~~&&\times[\Gamma_{Zuu}^{\sigma}S_q((x_2+x_3)p_2-y_3p_4)\Gamma_{guu}^{\rho}]_{\bar{\beta}\beta}T^a_{\bar{j}j}
[\Gamma_{gdd}^{\eta}]_{\bar{\kappa}\kappa}T^a_{\bar{k}k}\Phi_{P}^{\text{ini}}[i,j,k,\alpha,\beta,\kappa,p_2, \lambda_{2}] \nonumber\\
~~~~~~&& \times D_{g,\rho\eta}(y_3p_4-x_3p_2)D_{\gamma,\mu\omega}(y_1p_4-x_1p_2)D_{Z,\mu\sigma}((y_2+y_3)p_4-(x_2+x_3)p_2),
\label{eq:Amp-gammaZ-uu-1-a}
\end{eqnarray}
where the vertices are
\begin{eqnarray}
\Gamma_{Zee}^{\mu} &=& -i(\bar{g}_V^{e}\gamma^{\mu}+\bar{g}_A^{e}\gamma^{\mu}\gamma_{5}), \nonumber \\
\Gamma_{Zuu}^{\mu} &=& -i(\bar{g}_V^{u}\gamma^{\mu}+\bar{g}_A^{u}\gamma^{\mu}\gamma_{5}), \nonumber \\
\Gamma_{Zdd}^{\mu} &=& -i(\bar{g}_V^{d}\gamma^{\mu}+\bar{g}_A^{d}\gamma^{\mu}\gamma_{5}),
\label{eq:vertices-z}
\end{eqnarray}
with
\begin{eqnarray}
\bar{g}_{V,A}^{e,u,d} &=& -\frac{e}{4\sin\theta_W\cos\theta_W}g_{V,A}^{e,u,d},
\end{eqnarray}
and $\theta_W$ the Weinberg angle, and
\begin{eqnarray}
g_V^{e} &=& -1 + 4\sin^2\theta_W, \quad g_A^{e} = -1, \nonumber \\
g_V^{u} &=& 1 - 8/3\sin^2\theta_W, \quad g_A^{u} = 1, \nonumber \\
g_V^{d} &=& -1 + 4/3\sin^2\theta_W, \quad g_A^{d} = -1.
\label{eq:Amp-vertex-propagator-gVA}
\end{eqnarray}
The propagator of $Z$ boson is taken as
\begin{eqnarray}
D_{Z,\mu\nu}(k) \equiv (g_{\mu\nu}-(1-\xi_Z)\frac{k_{\mu}k_{\nu}}{k^2-\xi_ZM_Z^2+i\varepsilon})\frac{-i}{k^2-M_Z^2+i\varepsilon} = \frac{-i}{k^2-M_Z^2+i\varepsilon}g_{\mu\nu},
\end{eqnarray}
with $\xi_Z=1$.

\subsection{Invariant amplitudes}
To discuss the TBE contribution at the amplitude level, we express the amplitude in a general form as
\begin{eqnarray}
\mathcal{M} &=& \mathcal{M}^{\text{PC}}+\mathcal{M}^{\text{PV}},
\end{eqnarray}
where the indices $\text{PC}$ and $\text{PV}$ refer to the parity-conserving and parity-violating parts, respectively. In the practical calculation, we choose
\begin{eqnarray}
\mathcal{M}^{\text{PC,PV}} &=& \sum_{i=1}^{3} c_i^{\text{PC,PV}}\mathcal{M}_{i}^{\text{PC,PV}},
\end{eqnarray}
with
\begin{eqnarray}
\mathcal{M}_{1}^{\text{PC}} &\equiv& [\bar{u}_3\gamma_{\mu}u_1][\bar{u}_4\gamma^{\mu}u_2], \nonumber \\
\mathcal{M}_{2}^{\text{PC}} &\equiv& \frac{1}{Q}[\bar{u}_3\slashed{P}u_1][\bar{u}_4u_2], \nonumber \\
\mathcal{M}_{3}^{\text{PC}} &\equiv& \frac{1}{{Q^2}} [\bar{u}_3\slashed{P}u_1][\bar{u}_4\slashed{K}u_2],
\label{eq:invariant-Amp-TPE}
\end{eqnarray}
and
\begin{eqnarray}
\mathcal{M}_{1}^{\text{PV}} &\equiv& [\bar{u}_3\gamma_{\mu}\gamma_5u_1][\bar{u}_4\gamma^{\mu}u_2], \nonumber \\
\mathcal{M}_{2}^{\text{PV}} &\equiv& \frac{1}{Q}[\bar{u}_3\gamma_{\mu}\gamma_5u_1][\bar{u}_4i\sigma^{\mu\nu}q_{\nu}u_2], \nonumber \\
\mathcal{M}_{3}^{\text{PV}} &\equiv& \frac{1}{Q^2}[\bar{u}_3\slashed{P}\gamma_5u_1][\bar{u}_4\slashed{K}u_2].
\label{eq:invariant-Amp}
\end{eqnarray}
Here we note that, since we focus on the behavior of the amplitudes in the limit $M_N/Q\rightarrow0$, we use $Q$ (rather than $M_N$ as in \textcolor{black}{Refs.~\cite{pQCD-Kivel:2009eg,Guttmann:2013ndr}}) to scale the invariant amplitudes.

For the one-photon-exchange diagram, we have
\begin{eqnarray}
\mathcal{M}_{1\gamma} &=& -i\bar{u}_3(-ie\gamma^{\mu})u_1 \frac{-ig_{\mu\nu}}{(p_1-p_3)^2}\bar{u}_4\left[ie\left(F_1\gamma^{\nu}+F_2\frac{i\sigma^{\alpha\nu}}{2M_N}(p_4-p_2)_{\alpha}\right)\right]u_2,
\label{equation:amp-one-gamma}
\end{eqnarray}
where $F_1$ and $F_2$ are the Dirac and Pauli form factors. The amplitude is
\begin{eqnarray}
\mathcal{M}_{1\gamma} &\equiv& \sum_{i=1}^{3} c_i^{1\gamma} \mathcal{M}_{i}^{\text{PC}}.
\end{eqnarray}
Then, the corresponding coefficients are
\begin{eqnarray}
c_1^{1\gamma} = \frac{e^2(F_1+F_2)}{Q^2}, \quad c_2^{1\gamma} = \frac{-e^2F_2}{2M_NQ}, \quad c_3^{1\gamma} = 0.
\end{eqnarray}

In Ref.~\cite{pQCD-Kivel:2009eg}, the full amplitude for OPE + TPE diagrams is written as
\begin{eqnarray}
{\cal M} &=& \frac{e^2}{Q^2} \bar{u}_3 \gamma_\mu u_1 \times \bar{u}_4\left(\widetilde{G}_M \gamma^\mu-\widetilde{F}_2 \frac{P^\mu}{M_N}+\widetilde{F}_3 \frac{\gamma \cdot K P^\mu}{M_N^2}\right) u_2,
\end{eqnarray}
where
\begin{eqnarray}
\widetilde{G}_M &\equiv& G_M + \delta \widetilde{G}_M, \nonumber\\
\widetilde{F}_2 &\equiv& F_2 + \delta \widetilde{F}_2,
\end{eqnarray}
and
\begin{eqnarray}
G_M \equiv F_1 +F_2 , \quad  G_E\equiv F_1-\tau F_2.
\end{eqnarray}
Comparing with our notations, we have
\begin{eqnarray}
\frac{e^2}{Q^2}\delta \widetilde{G}_M &=& c_{1}^{(2\gamma)}, \nonumber\\
\frac{e^2}{M_NQ}\delta \widetilde{F}_2 &=& -c_{2}^{(2\gamma)}, \nonumber\\
\frac{e^2}{M_N^2}\widetilde{F}_3 &=&  c_{3}^{(2\gamma)}.
\end{eqnarray}

\subsection{Physical quantities expressed by $c_{i}$}
In principle, when $c_{i}^{1\gamma}$, $c_{i}^{1Z}$, $c_{i}^{2\gamma}$, and $c_{i}^{\gamma Z}$ are known, all OBE and TBE contributions to unpolarized and polarized observables are determined. In this subsection, we use these coefficients to express the unpolarized cross sections and $A_{\text{PV}}$.

For the unpolarized cross section, in the leading order of $\alpha_e$ with $\alpha_e\equiv e^2/(4\pi)$, one has
\begin{eqnarray}
\frac{d\sigma_{un}^{1\gamma\otimes1\gamma}}{d\Omega} &=& \frac{1}{64\pi^2s}\frac{E'}{E} \frac{1}{4} \sum_{\text {helicity }}\mathcal{M}_{1\gamma}\mathcal{M}_{1\gamma}^{*},
\end{eqnarray}
where $\mathcal{M}_{1\gamma}$ is given in Eq.~(\ref{equation:amp-one-gamma}), and $E'$ and $E$ are the energies of the initial and final electrons in the lab frame, respectively. This gives the usual form:
\begin{eqnarray}
\sigma_{R}^{1\gamma\otimes 1\gamma}\equiv \frac{1}{R}\frac{d\sigma_{un}^{1\gamma\otimes1\gamma}}{d\Omega} = G_{M}^2(Q^2) + \frac{\varepsilon}{\tau} G_{E}^2(Q^2),
\end{eqnarray}
where
\begin{eqnarray}
R\equiv \frac{1}{64\pi^2s} \frac{E'}{E} \frac{2e^4}{(1-\varepsilon)}, ~~~~~ \tau \equiv \frac{Q^2}{4M_N^2}.
\end{eqnarray}

For the TPE contribution, where the TPE amplitude is written as
\begin{eqnarray}
\mathcal{M}_{2\gamma} &\equiv&  \sum_{i=1}^{3} c_i^{2\gamma}\mathcal{M}_{i}^{\text{PC}},
\label{eq:amp-2gamma}
\end{eqnarray}
we have
\begin{eqnarray}
\frac{d\sigma_{un}^{1\gamma\otimes 2\gamma}}{d\Omega} = \frac{1}{64\pi^2s} \frac{E'}{E} \frac{1}{4} \sum_{\text {helicity }} 2\text{Re}[\mathcal{M}^{1\gamma\ast}\mathcal{M}_{2\gamma}],
\end{eqnarray}
which gives
\begin{eqnarray}
\sigma_{R}^{1\gamma\otimes2\gamma} \simeq 2G_M\left(\delta \widetilde{G}_M+\widetilde{F}_3 \varepsilon N \right)+\frac{2G_E\varepsilon}{\tau}\left(\delta \widetilde{G}_M +\widetilde{F}_3 N\right),
\end{eqnarray}
with
\begin{eqnarray}
N &\equiv& \frac{\sqrt{\left(1-\varepsilon^2\right) \tau(1+\tau)}}{1-\varepsilon}.
\end{eqnarray}
Here the contribution from  $\delta \widetilde{F}_2$ has been neglected since it is zero at the leading order of pQCD.

For the asymmetry quantity $A_{\text{PV}}$, which is defined as
\begin{eqnarray}
A_{\text{PV}} \equiv  \frac{\sum_{\text {helicity }}\left(\mathcal{M}_{+} \mathcal{M}_{+}^{*}-\mathcal{M}_{-} \mathcal{M}_{-}^{*}\right)}{\sum_{\text {helicity }}\left(\mathcal{M}_{+} \mathcal{M}_{+}^{*}+\mathcal{M}_{-} \mathcal{M}_{-}^{*}\right)},
\end{eqnarray}
where $\mathcal{M}_{+,-}$ are the helicity amplitudes for incoming-electron helicities $+,-$, respectively, the $\gamma Z$-exchange contribution can be written as
\begin{eqnarray}
A_{P V}^{\gamma \otimes \gamma Z} &=& \frac{1}{e^2\sigma} \sum_{i=1}^3 \mathcal{N}_i \operatorname{Re}[c_{i}^{\gamma Z,\text{PV}}]  \nonumber\\
 &=& \frac{1}{e^2\sigma} \sum_{i=1}^3 \mathcal{N}_i  \Big\{\operatorname{Re}[c_{i,V}^{\gamma Z,\text{PV}}] g_e^A+ \operatorname{Re}[c_{i,A}^{\gamma Z,\text{PV}}] g_e^V \Big\} \nonumber\\
~~~~~~& \equiv& \operatorname{Re}[\square_{\gamma Z}^{V}] + \operatorname{Re}[\square_{\gamma Z}^{A}],
\end{eqnarray}
where
\begin{eqnarray}
\sigma &=&4F_1^2M_N^2(16\nu^2-4M_N^2Q^2+Q^4)+16F_1F_2M_N^2Q^4+F_2^2Q^2(16\nu^2+4M_N^2Q^2-Q^4),
\label{equation:sigma}
\end{eqnarray}
and
\begin{eqnarray}
\mathcal{N}_1 &=& 8 M_N^2 Q^2\left[\left(16\nu^2-4 M_N^2 Q^2+Q^4\right) F_1+2 Q^4 F_2\right] \nonumber\\
\mathcal{N}_2 &=& 2 M_N Q^3\left[8 M_N^2 Q^2 F_1+\left(16\nu^2-4 M_N^2 Q^2+Q^4\right) F_2\right], \nonumber\\
\mathcal{N}_3 &=& 8 M_N^2 \nu\left(16\nu^2-4 M_N^2 Q^2+Q^4\right) F_1,
\end{eqnarray}
$c_{i,V}^{\gamma Z,\text{PV}}$, $c_{i,A}^{\gamma Z,\text{PV}}$, and $c_{i}^{\gamma Z,\text{PV}}$ are defined as
\begin{eqnarray}
\mathcal{M}_{\gamma Z} &\equiv&  \sum_{i=1}^{3}(c_i^{\gamma Z,\text{PC}}\mathcal{M}_{i}^{\text{PC}}+c_i^{\gamma Z,\text{PV}}\mathcal{M}_{i}^{\text{PV}}),\nonumber\\
c_i^{\gamma Z,\text{PV}} &\equiv& c_{i,V}^{\gamma Z,\text{PV}}g_{e}^{A} + c_{i,A}^{\gamma Z,\text{PV}}g_{e}^{V}.
\label{eq:amp-gamma-Z}
\end{eqnarray}
Here we note that the definitions of $P$, $K$, $\nu$, and $c_i$ are slightly different from those in Ref.~\cite{QianQianGuo2023}.

\section{Analytical results}

In the large-$Q^2$ limit, where we approximately take $m_e\rightarrow0$ and $M_N/Q\rightarrow0$, the TPE contribution from Fig.~\ref{Figure:ep-ep-TPE-uu-a} to the coefficients is
\begin{eqnarray}
\delta \widetilde{G}_M^{(a)} &=& \int^{1}_{0} [dxdy] \frac{-e^2Q_u^2g_s^2s[4 T T'+(A-V)(A'-V')]}{36Q^4x_3(x_2+x_3)^2 y_1y_3(y_2+y_3) \left(s(x_1-y_1)+Q^2(x_2+x_3) y_1 +i\varepsilon\right)},  \nonumber \\
\delta \widetilde{F}^{(a)}_2&=&0, \nonumber \\
\frac{\nu}{M_N^2}\widetilde{F}_3^{(a)} &=& \int^{1}_{0} dx_idy_i \frac{-e^2Q_u^2g_s^2s[4 T T'+(A-V)(A'-V')]}{18Q^4x_1x_3(x_2+x_3) y_1y_3(y_2+y_3) \left(s(x_1-y_1)+Q^2(x_2+x_3) y_1 +i\varepsilon\right)},
\label{equation:F3-Gm-Fig-1-a}
\end{eqnarray}
where the variables in $V,T,A$ and $V',T',A'$ are $(x_1,x_2,x_3)$ and $(y_1,y_2,y_3)$, respectively.

Similarly, the coefficients for all Feynman diagrams are
\begin{eqnarray}
\delta \widetilde{G}_M &=&  -\frac{\alpha_{em}\alpha_{s}}{Q^4}\frac{16\pi^2}{9}(2\zeta-1)\int^{1}_{0} [dxdy] \frac{x_2y_2}{D}\{Q_u^2[4TT'+(V+A)(V'+A')](3,2,1)\nonumber\\
~~~~~~&& + Q_uQ_d[4TT'+(V+A)(V'+A')](1,2,3)+2Q_uQ_d[VV'+AA'](1,3,2)\},  \nonumber \\
\delta \widetilde{F}_2&=&0, \nonumber \\
\frac{\nu}{M_N^2}\widetilde{F}_3 &=& -\frac{\alpha_{em}\alpha_{s}}{Q^4}\frac{8\pi^2}{9}(2\zeta-1)\int^{1}_{0} [dxdy] \frac{(x_2\bar{y}_2+\bar{x}_2y_2)}{D}\{Q_u^2[4TT'+(V+A)(V'+A')](3,2,1)\nonumber\\
~~~~~~&& + Q_uQ_d[4TT'+(V+A)(V'+A')](1,2,3)+2Q_uQ_d[VV'+AA'](1,3,2)\},
\label{equation:Gm-F3-finally}
\end{eqnarray}
where some variable transformations are used. The symbol $(1,2,3)$ means that the variables in $V,A,T$ and $V',A',T'$ are $x_{1,2,3}$ and $y_{1,2,3}$ in order, $\zeta\equiv s/Q^2$ with $\zeta\geq1$, and $\bar{\zeta}\equiv1-\zeta$. We define
\begin{eqnarray}
D &=& (y_1y_2\bar{y}_2)(x_1x_2\bar{x}_2)[x_2\bar{\zeta}+y_2\zeta-x_2y_2+i\varepsilon] [x_2\zeta+y_2\bar{\zeta}-x_2y_2+i\varepsilon]
\label{equation:D}.
\end{eqnarray}
The above results are the same as those given in Ref.~\cite{pQCD-Kivel:2009eg}. In this work, we also calculate the $\gamma Z$-exchange contribution to the corresponding PV coefficients, which are

\begin{eqnarray}
c_{1,A}^{\gamma Z,\text{PV}} &=& \frac{16\pi^3 \alpha_{e}^2 \alpha_s }{9Q^4\sin^22\theta_W} \int^{1}_{0} [dxdy] (x_2y_2-2(x_2+y_2)\zeta\bar{\zeta})[ M_{(3,2,1)}^{A} - M_{(1,2,3)}^{A} + M_{(1,3,2)}^{A}],  \nonumber \\
c_{3,A}^{\gamma Z,\text{PV}} &=& \frac{32\pi^3 \alpha_{e}^2 \alpha_s }{9Q^4\sin^22\theta_W} \int^{1}_{0} [dxdy] (x_2+y_2)(2\zeta-1)[ -M_{(3,2,1)}^{A} + M_{(1,2,3)}^{A} - M_{(1,3,2)}^{A}],\nonumber \\
c_{1,V}^{\gamma Z,\text{PV}} &=& \frac{16\pi^3 \alpha_{e}^2 \alpha_s}{9Q^4\sin^22\theta_W} \int^{1}_{0} [dxdy] x_2y_2(2\zeta-1)[ M_{(3,2,1)}^{V} +  M_{(1,2,3)}^{V} + M_{(1,3,2)}^{V}],\nonumber \\
c_{3,V}^{\gamma Z,\text{PV}} &=& \frac{32\pi^3 \alpha_{e}^2 \alpha_s }{9Q^4\sin^22\theta_W} \int^{1}_{0} [dxdy] (\bar{x}_2y_2+x_2\bar{y}_2)[ M_{(3,2,1)}^{V} + M_{(1,2,3)}^{V} + M_{(1,3,2)}^{V}],
\label{eq:c1-c3}
\end{eqnarray}
with
\begin{eqnarray}
M_{(3,2,1)}^{A} &=& Q_u g_A^u\Big[\frac{4TT'-(A+V)(A'+V')}{D_{1}}+\frac{4TT'+(A+V)(A'+V')}{D_{2}}\Big], \nonumber \\
M_{(1,2,3)}^{A} &=& \frac{Q_u g_A^d[4TT'-(A+V)(A'+V')]}{D_{1}}-\frac{Q_d g_A^u[4TT'+(A+V)(A'+V')]}{D_{2}}, \nonumber \\
M_{(1,3,2)}^{A} &=& \frac{2Q_d g_A^u[A'V+AV']}{D_{1}}+\frac{2Q_u g_A^d[AA'+VV']}{D_{2}},\nonumber\\
M_{(3,2,1)}^{V} &=& Q_u g_V^u[\frac{1}{D_{1}}+\frac{1}{D_{2}}][4TT'+(A+V)(A'+V')], \nonumber \\
M_{(1,2,3)}^{V} &=& [\frac{Q_u g_V^d}{D_{1}}+\frac{Q_d g_V^u}{D_{2}}][4TT'+(A+V)(A'+V')], \nonumber \\
M_{(1,3,2)}^{V} &=& [\frac{2Q_d g_V^u}{D_{1}}+\frac{2Q_u g_V^d}{D_{2}}][AA'+VV'],
\label{eq:M-D}
\end{eqnarray}
 and
\begin{eqnarray}
D_{1} &=& x_1x_2y_1y_2(M_Z^2+Q^2\bar{x}_2\bar{y}_2)(x_2\zeta+y_2\bar{\zeta}-x_2y_2+i\epsilon)(x_2\bar{\zeta}+y_2\zeta-x_2y_2+i\epsilon), \nonumber \\
D_{2} &=& x_1\bar{x}_2y_1\bar{y}_2(M_Z^2+Q^2x_2y_2)(x_2\zeta+y_2\bar{\zeta}-x_2y_2+i\epsilon)(x_2\bar{\zeta}+y_2\zeta-x_2y_2+i\epsilon).
\end{eqnarray}
where labels such as $(1,2,3)$ denote that the variables in $V,T,A$ and $V',T',A'$ are $(x_1,x_2,x_3)$ and $(y_1,y_2,y_3)$, respectively.

We note that a similar $\gamma Z$-exchange contribution in pQCD is also calculated in the PhD thesis \cite{Guttmann:2013ndr}. \textcolor{black}{Ref.~\cite{Guttmann:2013ndr} expresses the $\gamma Z$ box in terms of effective form-factor-like quantities such as $\delta\widetilde{G}_M^Z$, whereas we work directly with the invariant amplitudes $c_{i,V/A}^{\gamma Z,\text{PV}}$ defined in Eq.~(\ref{eq:amp-gamma-Z}). As one explicit example, the $Q_u g_V^d$ term in $\delta\widetilde{G}_M^Z$ of Ref.~\cite{Guttmann:2013ndr} contains the operator structure $(V'-A')(V-A)+4T'T$ with the quark-label ordering $(1,3,2)$, whereas the corresponding term in our $c_{1,V}^{\gamma Z,\text{PV}}$ contains $(V'+A')(V+A)+4T'T$ with ordering $(3,2,1)$ or $(V'-A')(V-A)+4T'T$ with ordering $(3,1,2)$. Similar differences appear in the relative signs and in the quark permutations of the other terms. In addition, our analytical expressions retain the full $Z$-boson mass dependence through the propagator denominators $D_1$ and $D_2$, whereas the comparison formula in Ref.~\cite{Guttmann:2013ndr} is organized in the high-$Q^2$ limit. These differences reflect different amplitude decompositions and conventions, not merely a change of notation. Our results reduce to the same high-$Q^2$ scaling.}

\section{Numerical results}

To numerically calculate the integrations in Eqs.~(\ref{equation:Gm-F3-finally},~\ref{eq:c1-c3}), one can reduce the denominators by applying the change $x_1\leftrightarrow x_2$, $y_1\leftrightarrow y_2$, and then integrate $x_3,y_3,x_1,y_1$ analytically after substituting the functions $V,T,A$. For the remaining integrals over $x_2, y_2$, there is a pole in $y_2$ in the physical region:
\begin{eqnarray}
y_2^{pole}=\frac{sx_2}{s + Q^2 (-1 + x_2)}+i\epsilon.
\end{eqnarray}
After the pole position is known, the integrals can be done directly.

\subsection{TPE contributions to unpolarized cross section}

\begin{figure}[!htbp]
\begin{center}
\includegraphics[height=10.0cm]{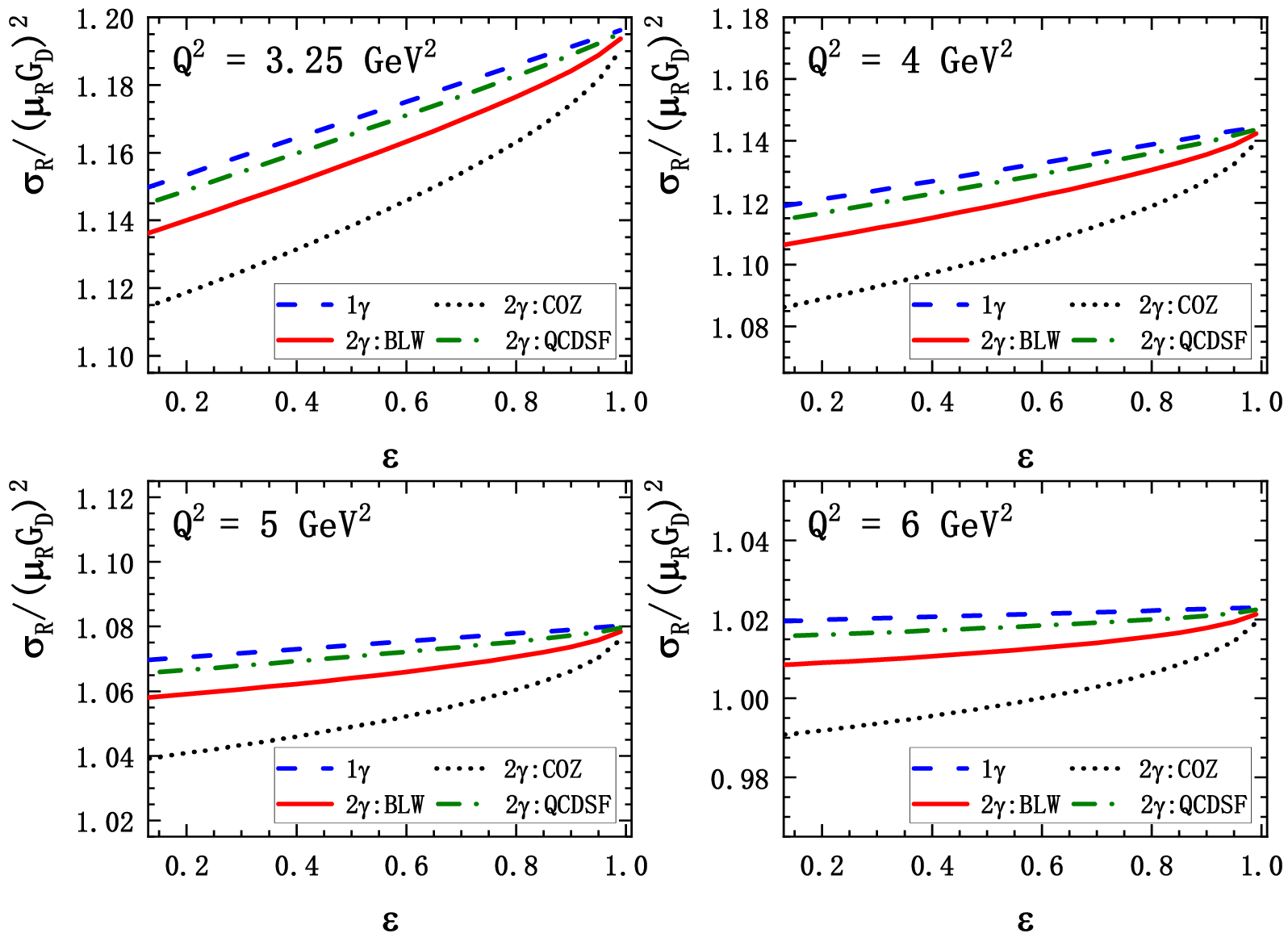}
\caption{Numerical results for $\sigma_{R}/(\mu_pG_D)^2$. The dashed blue curves refer to the $1\gamma$-exchange contribution. The dotted black, solid red, and dash-dotted olive curves refer to results with the COZ, BLW, and QCDSF models, respectively.}
\label{Figure:cross-section}
\end{center}
\end{figure}

For comparison, we present the TPE contributions to $\sigma_{R}^{1\gamma\otimes1\gamma}/(\mu_{p}G_D)^2$ and $\sigma_{R}^{1\gamma\otimes2\gamma}/(\mu_{p}G_D)^2$ as functions of $\epsilon$ in Fig.~\ref{Figure:cross-section}, where
\begin{eqnarray}
G_D\equiv \frac{1}{(1+Q^2/0.71)^2},
\end{eqnarray}
Here $\mu_p=2.79$ is the proton magnetic moment, and
\begin{eqnarray}
\epsilon &=& [1 + 2(1 + \tau \tan^2 (\theta_{e}/2))]^{-1},\nonumber\\
\tau &=&\frac{Q^2}{4M_N^2},
\end{eqnarray}
with $\theta_e$ the scattering angle in the lab frame.

Although the analytical expressions are equivalent to those in Ref.~\cite{pQCD-Kivel:2009eg}, our numerical results are slightly different. This difference may come from the values of $G_{E},G_{M}$ used in $\sigma_{R}^{1\gamma\otimes1\gamma}$. In our calculation, we take $G_{E}, G_{M}$ from Refs.~\cite{GE-GM-ratio-JeffersonLabHallA:1999epl,GE-GM-ratio-JeffersonLabHallA:2001qqe,GE-GM-ratio-Punjabi:2005wq}. As a validation benchmark, our TPE numerical implementation still reproduces the expected magnitude and kinematic trend of the known pQCD contribution to the reduced unpolarized cross section.

\subsection{$\gamma Z$-exchange contributions $c_{i,V}^{\gamma Z\text{PV}}$ and $c_{i,A}^{\gamma Z\text{PV}}$ }

\begin{figure}[!htbp]
\begin{center}
\includegraphics[height=10.0cm]{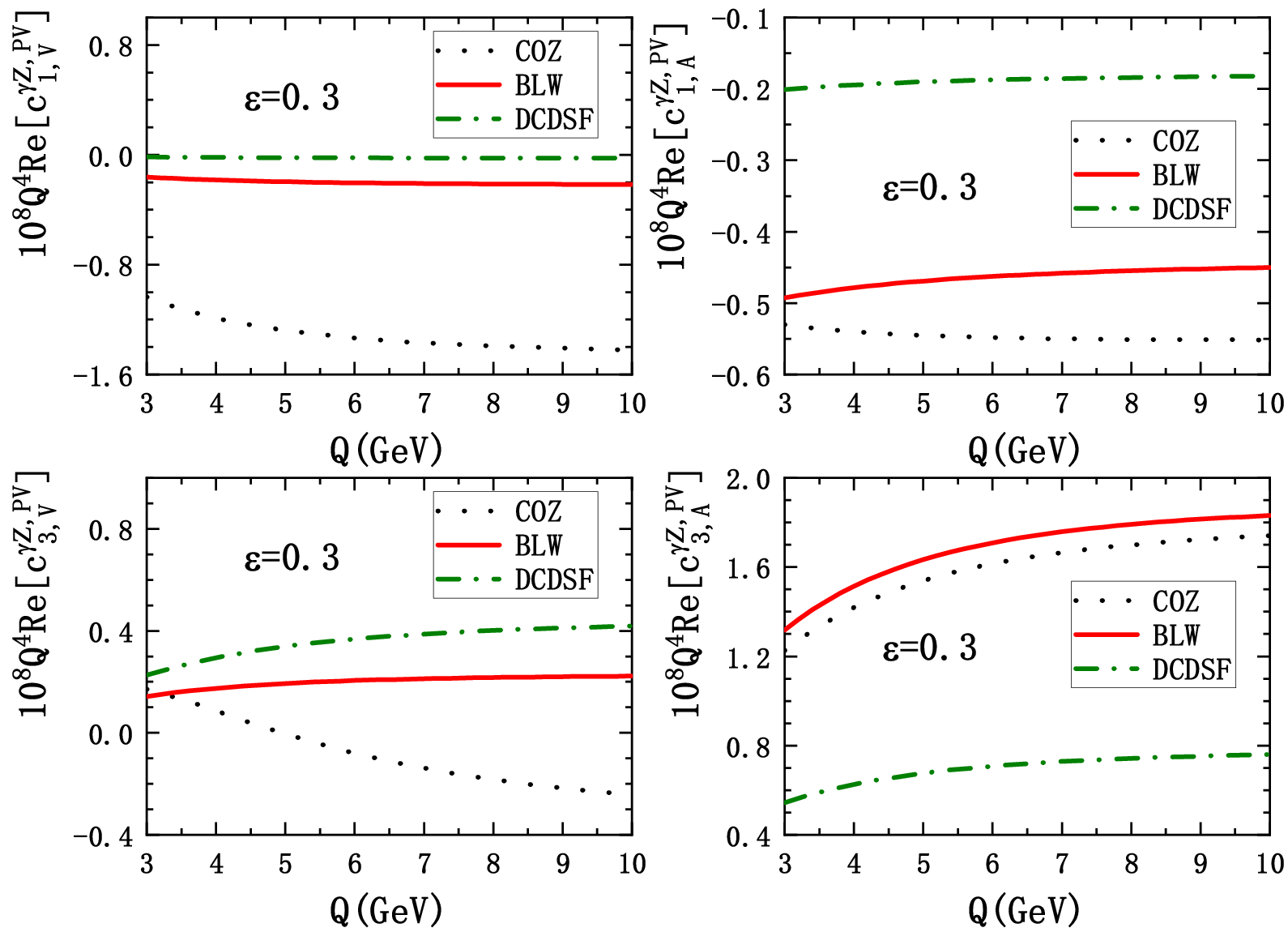}
\caption{Numerical results for $\text{Re}[c_{1,V}^{\gamma Z,\text{PV}}]$, $\text{Re}[c_{3,V}^{\gamma Z,\text{PV}}]$, $\text{Re}[c_{1,A}^{\gamma Z,\text{PV}}]$, and $\text{Re}[c_{3,A}^{\gamma Z,\text{PV}}]$ as functions of $Q$ at $\varepsilon=0.3$. The dotted black, solid red, and dash-dotted olive curves refer to results with the COZ, BLW, and QCDSF models, respectively.}
\label{Figure:ci-vs-Q}
\end{center}
\end{figure}

The numerical results for the coefficients $c_{i,V}^{\gamma Z,\text{PV}}$ and $c_{i,A}^{\gamma Z,\text{PV}}$ as functions of $Q^2$ are presented in Fig.~\ref{Figure:ci-vs-Q}, where we take $\varepsilon = 0.3$ as an example and show only the real parts. The dotted black, solid red, and dash-dotted olive curves refer to results using the COZ \cite{COZ-definition}, BLW \cite{BLW-definition}, and QCDSF \cite{QCDSF-definition} models, respectively. Fig.~\ref{Figure:ci-vs-Q} provides numerical evidence for this asymptotic behavior:
\begin{eqnarray}
c_{i,A}^{\gamma Z,\text{PV}} \sim c_{1,V}^{\gamma Z,\text{PV}} \sim Q^{-4}.
\end{eqnarray}

\textcolor{black}{Having examined the $Q^2$ dependence in Fig.~\ref{Figure:ci-vs-Q}, we now discuss the fixed-$Q^2$ DRs for the amplitude coefficients at high energy. At low energy, DRs for the combinations $\square_{\gamma Z}^{V,A}$ that enter $A_{\text{PV}}$ are widely used \cite{DRs-Gorchtein:2008px,DRs-Sibirtsev:2010zg,DRs-Gorchtein:2011mz,DRs-Erler:2019rmr}. For example, DRs of the following form are used:
\begin{eqnarray}
\textrm{Re}[X_{\gamma Z}(\nu,Q^2)] = \frac{2 \nu}{\pi} \mathrm{P}\Big[\int_{\nu_{th}}^{\infty} \frac{\textrm{Im}[X_{\gamma Z}(\bar{\nu}^{+},Q^2)\big]}{\bar{\nu}^{2}-\nu^{2}} d \bar{\nu}\Big], \nonumber\\
\textrm{Re}[Y_{\gamma Z}(\nu,Q^2)] = \frac{2 }{\pi} \mathrm{P}\Big[\int_{\nu_{th}}^{\infty} \frac{\bar{\nu}\textrm{Im}[Y_{\gamma Z}(\bar{\nu}^{+},Q^2)\big]}{\bar{\nu}^{2}-\nu^{2}} d \bar{\nu}\Big],
\label{equation:DR3}
\end{eqnarray}
where $\bar{\nu}^{+}\equiv \bar{\nu}+i\epsilon$ denotes the integration variable taken on the physical sheet, $\mathrm{P}$ denotes the Cauchy principal value of the integral, $X$ refers to $\square_{\gamma Z}^{V}$, and $Y$ refers to $\square_{\gamma Z}^{A}$. Ref.~\cite{QianQianGuo2024} shows that, at finite $Q^2$, the appropriate DRs are those for $c_i^{\gamma Z,\text{PV}}$ rather than for $\square_{\gamma Z}^{V,A}$, which are reliable only in the forward approximation. At the P2 design kinematics ($Q^2 = 0.0045$~GeV$^2$, $E = 0.155$~GeV), for example, the DR written directly for $\square_{\gamma Z}^{V}$ is significantly affected, whereas that for $c_i^{\gamma Z,\text{PV}}$ remains reliable. Accordingly, the same DR form in Eq.~(\ref{equation:DR3}) applies with $X=c_1^{\gamma Z,\text{PV}}$ and $Y=c_3^{\gamma Z,\text{PV}}$, and these DRs remain reliable at finite $Q^2$.}

The high-$\nu$ asymptotic behavior of the amplitudes determines the convergence of the DR integrals, and the convergence properties determine the required subtraction order. For the $X$-type DR, the large-$\bar{\nu}$ kernel behaves as $\sim 1/\bar{\nu}^{2}$ (at fixed external $\nu$), so convergence is controlled by the integral of  $\text{Im}X_{\gamma Z}(\bar{\nu})/\bar{\nu}^{2}$. For the $Y$-type DR, the kernel behaves as $\sim 1/\bar{\nu}$, so convergence is controlled by the integral of $\text{Im}Y_{\gamma Z}(\bar{\nu})/\bar{\nu}$. Therefore, if $\text{Im}Y_{\gamma Z}$ approaches a constant at high energy, the unsubtracted $Y$-type DR develops a logarithmic ultraviolet divergence and a subtraction is required. In contrast, quantities with sufficiently suppressed high-$\nu$ behavior remain compatible with the unsubtracted form.

\textcolor{black}{Since the physical DRs should be the same at low and high energy, we test the DRs in Eq.~(\ref{equation:DR3}) for $c_i^{\gamma Z,\text{PV}}$ in the high-$Q^2$ regime. The present paper complements the low-energy analysis of Ref.~\cite{QianQianGuo2024} by using pQCD to determine the high-$\nu$ behavior of $c_i^{\gamma Z,\text{PV}}$ and thereby the subtraction order required beyond the forward limit.}

\begin{figure}[!htbp]
\centering
\includegraphics[height = 10.0cm]{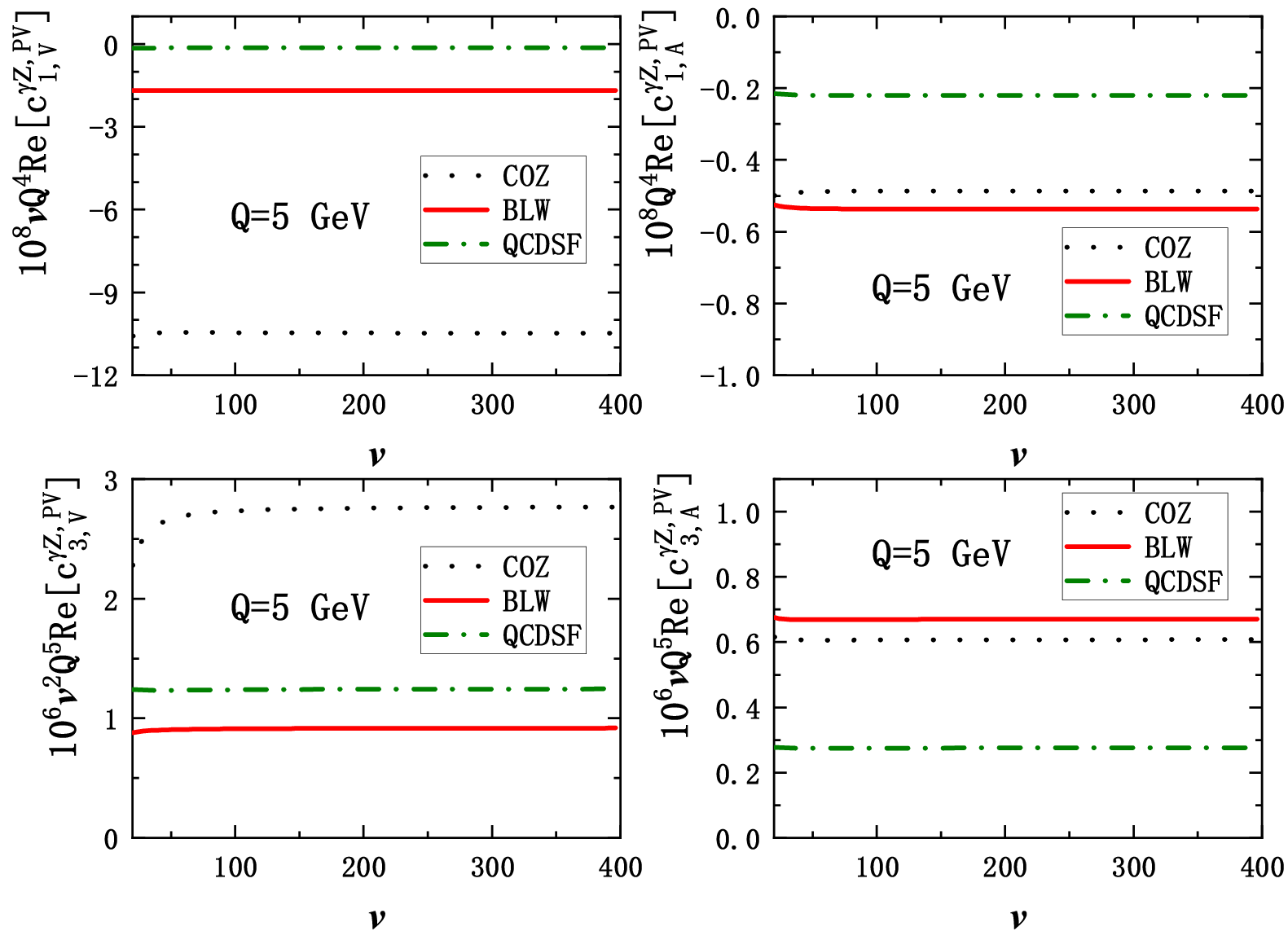}
\caption{Numerical results for $\text{Re}[c_{i,V}^{\gamma Z,\text{PV}}]$ and $\text{Re}[c_{i,A}^{\gamma Z,\text{PV}}]$ as functions of $\nu$ at $Q=5$ GeV. The dotted black, solid red, and dash-dotted olive curves refer to results obtained with the COZ, BLW, and QCDSF models, respectively.}
\label{Figure:ci-nu-Q5-Re}
\end{figure}

\begin{figure}[!htbp]
\centering
\includegraphics[height = 10.0cm]{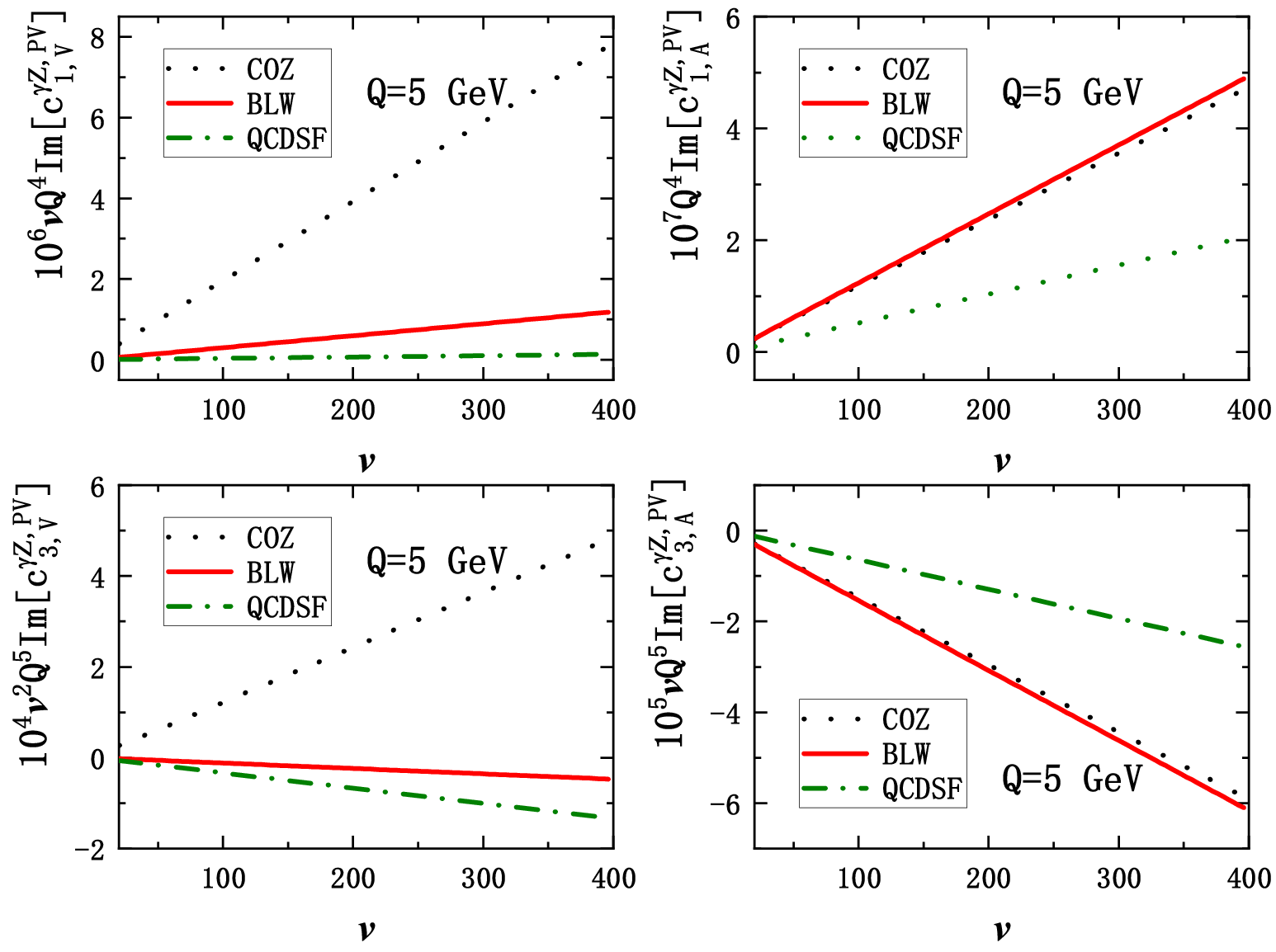}
\caption{Numerical results for $\text{Im}[c_{i,V}^{\gamma Z,\text{PV}}]$ and $\text{Im}[c_{i,A}^{\gamma Z,\text{PV}}]$ as functions of $\nu$ at $Q=5$ GeV. The dotted black, solid red, and dash-dotted olive curves refer to results obtained with the COZ, BLW, and QCDSF models, respectively.}
\label{Figure:ci-nu-Q5-Im}
\end{figure}

The high-$\nu$ behavior indicates that not all quantities satisfy the commonly used unsubtracted DR form. In Fig.~\ref{Figure:ci-nu-Q5-Re} and Fig.~\ref{Figure:ci-nu-Q5-Im}, we present the numerical results for the coefficients $c_{i,V}^{\gamma Z,\text{PV}}$ and $c_{i,A}^{\gamma Z,\text{PV}}$ as functions of $\nu$ at $Q=5$ GeV. The dotted black, solid red, and dash-dotted olive curves refer to results obtained with the COZ, BLW, and QCDSF models, respectively.

The results clearly show the high-energy behavior when $\nu\rightarrow\infty$:
\begin{eqnarray}
\text{Re}[c_{1,A}^{\gamma Z,\text{PV}}] &\sim&  \nu^{0} ,\nonumber\\
\text{Re}[c_{1,V}^{\gamma Z,\text{PV}}] &\sim& \text{Re}[c_{3,A}^{\gamma Z,\text{PV}}] \sim \nu^{-1},\nonumber\\
\text{Re}[c_{3,V}^{\gamma Z,\text{PV}}] &\sim& \nu^{-2},
\end{eqnarray}
and
\begin{eqnarray}
\text{Im}[c_{1,A}^{\gamma Z,\text{PV}}] &\sim&  \nu ,\nonumber\\
\text{Im}[c_{1,V}^{\gamma Z,\text{PV}}] &\sim& \text{Im}[c_{3,A}^{\gamma Z,\text{PV}}] \sim \nu^{0},\nonumber\\
\text{Im}[c_{3,V}^{\gamma Z,\text{PV}}] &\sim& \nu^{-1}.
\end{eqnarray}

These behaviors mean that, at high $Q^2$, $c_{1,A}^{\gamma Z,\text{PV}}$ and $c_{3,A}^{\gamma Z,\text{PV}}$ do not satisfy the corresponding DRs in Eq.~(\ref{equation:DR3}), and subtraction terms should be added.

\subsection{$\gamma Z$-exchange contributions to $\square_{\gamma Z}^{V,A}$}

\begin{figure}[!htbp]
\centering
\includegraphics[height = 5.0cm]{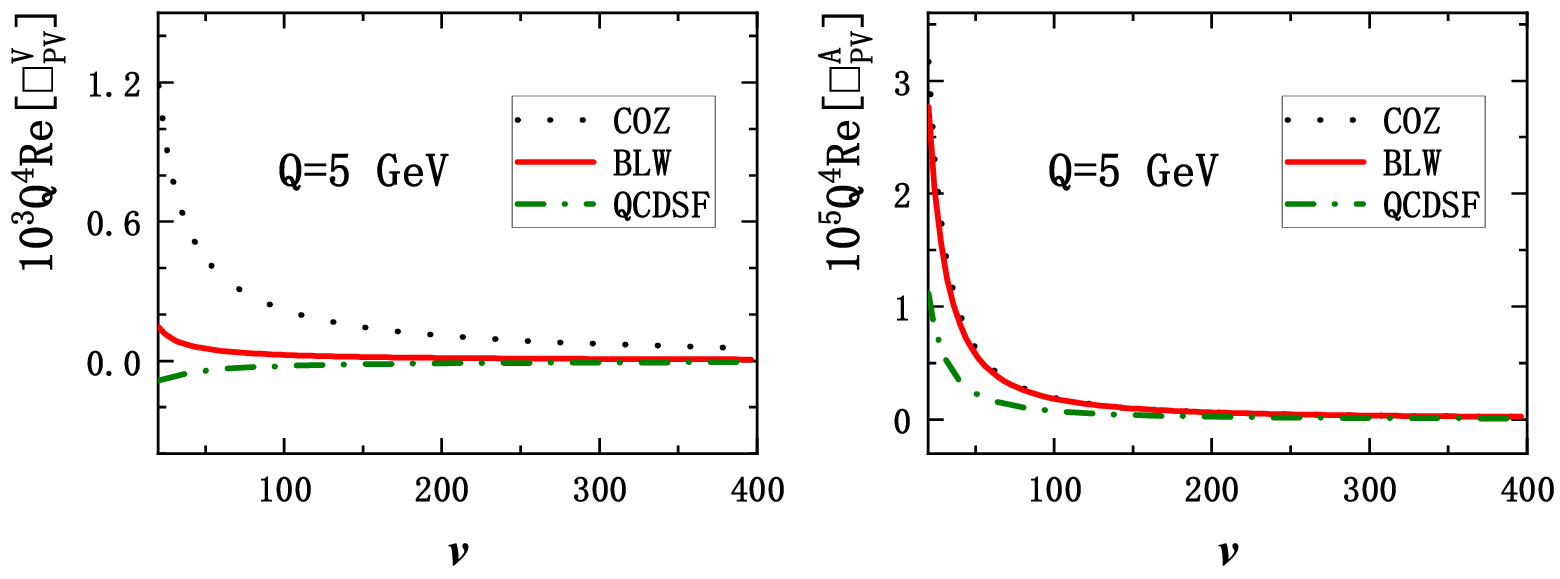}
\caption{Numerical results for $\text{Re}[\square_{\gamma Z}^{V,A}]$ as functions of $\nu$ at $Q=5$ GeV. The dotted black, solid red, and dash-dotted olive curves refer to results with the COZ, BLW, and QCDSF models, respectively.}
\label{Figure:Apv-nu-Q5-Re}
\end{figure}

\begin{figure}[!htbp]
\centering
\includegraphics[height = 5.0cm]{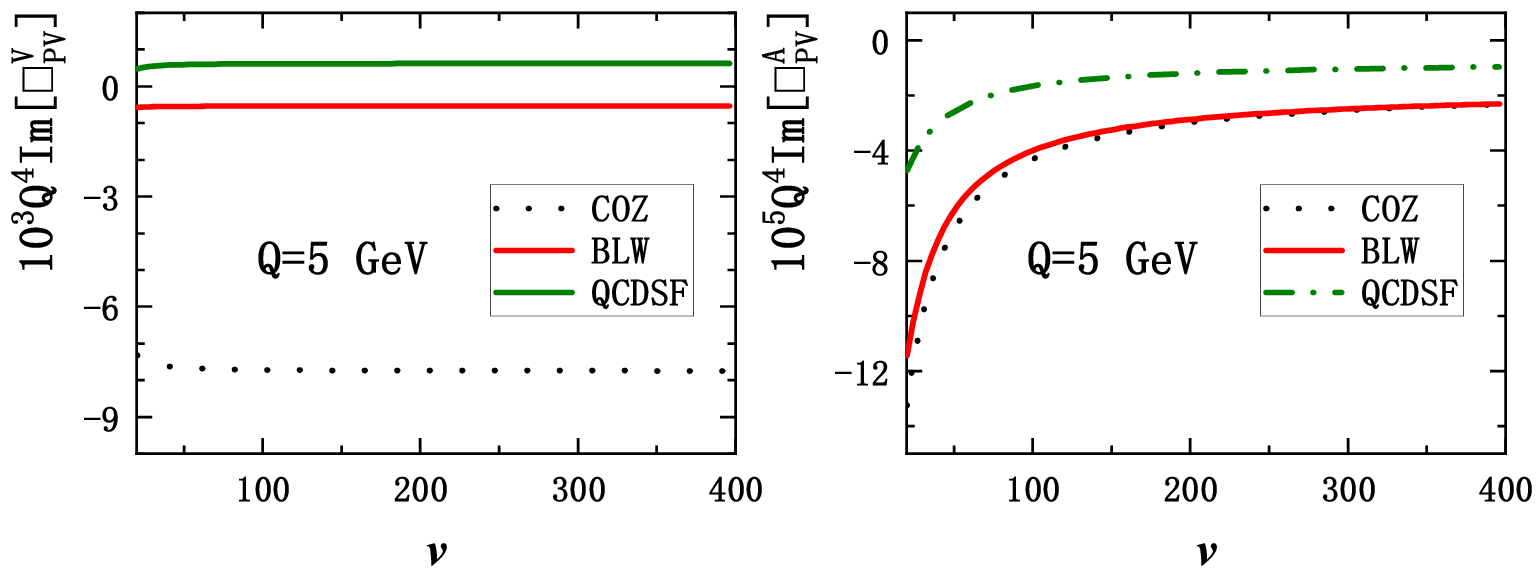}
\caption{Numerical results for $\text{Im}[\square_{\gamma Z}^{V,A}]$ as functions of $\nu$ at $Q=5$ GeV. The dotted black, solid red, and dash-dotted olive curves refer to results with the COZ, BLW, and QCDSF models, respectively.}
\label{Figure:Apv-nu-Q5-Im}
\end{figure}

\textcolor{black}{For completeness, Figs.~\ref{Figure:Apv-nu-Q5-Re} and~\ref{Figure:Apv-nu-Q5-Im} also show the high-$\nu$ behavior of $\square_{\gamma Z}^{V,A}$, but the DR analysis in this section remains centered on $c_i^{\gamma Z,\text{PV}}$.}

At low $Q^2$, unsubtracted DRs similar to Eq.~(\ref{equation:DR3}) are usually written for $\square_{\gamma Z}^{V,A}$. As discussed above, Ref.~\cite{QianQianGuo2024} shows that such DRs are reliable only in the forward approximation.

These DRs can also be tested in the high-energy case using the above results. We present the high-$\nu$ behaviors for $\square_{\gamma Z}^{V,A}$ in Fig.~\ref{Figure:Apv-nu-Q5-Re} and Fig.~\ref{Figure:Apv-nu-Q5-Im}, where we take $Q=5$ GeV as an example. Here, $F_1$ and $F_2$ are approximated as
\begin{eqnarray}
F_1 &\approx& \frac{1+\mu_p\tau}{(1+\tau)}G_D ,\nonumber\\
F_2 &\approx& \frac{\mu_p-1}{(1+\tau)}G_D.
\end{eqnarray}

The results in Fig.~\ref{Figure:Apv-nu-Q5-Re} and Fig.~\ref{Figure:Apv-nu-Q5-Im} clearly show that when $\nu\rightarrow\infty$, we have
\begin{eqnarray}
\text{Re}[\square_{\gamma Z}^{V}] &\sim& \text{Re}[\square_{\gamma Z}^{A}] \sim 0, \nonumber\\
\text{Im}[\square_{\gamma Z}^{V}] &\sim& \text{Im}[\square_{\gamma Z}^{A}] \sim \nu^{0},
\end{eqnarray}
These behaviors mean that \textcolor{black}{the DR in Eq.~(\ref{equation:DR3}) applied directly to $\square_{\gamma Z}^{A}$ is shown in Figs.~\ref{Figure:Apv-nu-Q5-Re} and~\ref{Figure:Apv-nu-Q5-Im} for illustration only. The subtracted form required at high $Q^2$ applies to $c_{3,A}^{\gamma Z,\text{PV}}$, as discussed above.}

\section{Summary}

In this work, we calculate the $\gamma Z$-exchange contribution at the amplitude level in elastic $ep$ scattering within pQCD. At fixed $Q^2$, the high-$\nu$ asymptotic behavior of the coefficients $c_i^{\gamma Z,\text{PV}}$ provides a direct link between pQCD scaling and the convergence properties of the DR integrals, thereby determining the required subtraction order in the regime beyond the forward limit ($s/Q^2\to\infty$). \textcolor{black}{Within the present pQCD framework and adopted distribution amplitudes, we find that $c_{3,A}^{\gamma Z,\text{PV}}$ requires a once-subtracted DR at high $Q^2$. Following Ref.~\cite{QianQianGuo2024}, the DRs are applied to the amplitude coefficients rather than directly to $\square_{\gamma Z}^{V,A}$, which are reliable only in the forward approximation.}

\section{Acknowledgments}

H.~Q.~Zhou would like to thank Zhi-Hui Guo for helpful discussions. H.~Q.~Zhou is supported by the National Natural Science Foundation of China under Grants Nos.~12150013 and 12075058. H.~Y.~C is supported by the Science and Technology Research Project of Hubei Provincial Education Department under Grant No.~20222502, the Hubei Provincial Natural Science Foundation under Grant No.~2023AFB443, and the National Natural Science Foundation of China under Grant No.~12405153.

\end{document}